\documentstyle[psfig,manuscript,aps]{revtex}
\tightenlines
\begin{document}

\title
{Screened Coulomb interactions in metallic alloys: I. Universal
screening in the atomic sphere approximation}

\author{A.~V. Ruban and H.~L. Skriver}

\address{
Center for Atomic-scale Materials Physics and Department of Physics, \\
Technical University of Denmark, DK-2800 Lyngby, Denmark}

\date{29 May 2001}

\maketitle
\begin{abstract}
We have used the locally self-consistent Green's function (LSGF) method in
supercell calculations to establish the distribution of the net charges
assigned to the atomic spheres of the alloy components in metallic alloys
with different compositions and degrees of order. This allows us to determine
the Madelung potential energy of a random alloy in the single-site mean field
approximation which makes the conventional single-site density-functional-
theory coherent potential approximation (SS-DFT-CPA) method practically 
identical to the supercell LSGF method with a single-site local interaction 
zone that yields an exact solution of the DFT problem. We demonstrate that 
the basic mechanism which governs the charge distribution is the screening 
of the net charges of the alloy components that makes the direct Coulomb 
interactions short-ranged. In the atomic sphere approximation, this screening 
appears to be almost independent of the alloy composition, lattice spacing, 
and crystal structure. A formalism which allows a consistent treatment of the
screened Coulomb interactions within the single-site mean-filed approximation 
is outlined. We also derive the contribution of the screened Coulomb 
interactions to the $S^{(2)}$ formalism and the generalized perturbation 
method.
\end{abstract}
\vspace{10mm}
\narrowtext

\section{Introduction}

The coherent potential approximation (CPA) \cite{soven67,taylor67,%
kirkpatrick70} as implemented on the basis of multiple scattering theory
\cite{gyorffy78,faulkner82} and combined with density functional theory
(DFT) \cite{hohenberg64,kohn65,johnson86,johnson90} constitutes the basis
for {\it ab inito\/} calculations of the electronic structure and physical
properties of random metallic alloys. This combination of the CPA with DFT, 
or, in most cases, with the local density approximation (LDA), seems to be 
quite transparent \cite{johnson86,johnson90} leading to expressions for the 
one-electron potential and total energy which are very similar to those for 
ordered systems. However, there is, by now, a well-recognized problem 
\cite{magri90,abrikosov92,johnson93,singh93} with this description caused by 
the fact that the atomic or "muffin-tin" spheres, which {\it artificially\/} 
divide the crystal into regions associated with particular alloy components, 
may poses non-zero net charges.

The problem stems from the fact that the conventional single-site (SS)
DFT-CPA method is based on the effective medium model of a random alloy which 
considers only conditionally averaged quantities and leads to the use of the 
single-site approximation not only in the electronic structure part of the 
problem during the solution of the CPA equations, but also in the DFT 
self-consistent loop in the calculations of the electrostatic contributions 
to the one-electron potential and energy. The single-site approximation 
provides no information as to the charge distribution beyond the atomic 
sphere of each alloy component and, since the surrounding effective medium is 
electroneutral, Poisson's equation cannot be solved properly if the atomic 
spheres have non-zero net charges. Hence, to find the correct solution to
Poisson's equation one must somehow describe the effect of the missing 
charge. Since the electron density inside each atomic sphere is well-defined,
any such description may be associated with a modification of the effective 
medium specifically for {\it each\/} alloy component. This may be regarded as
an inconsistency since, in that case, the CPA and the electrostatic part of 
the DFT are based on different effective media.

One obvious solution to the problem is to use electroneutral spheres (see, 
for instance, Ref. \cite{singh93}). However, in the methods based on the 
atomic sphere approximation (ASA) this frequently leads to large sphere 
overlaps and a quite poor description of the electronic structure, especially 
in the case of inhomogeneous system, such as partially ordered alloys or 
surfaces with an inhomogeneous concentration profile.

A more general solution can be found, however, in which the electrostatic 
potential is modified without making effective media for each alloy component
in contradiction to the assumptions of the CPA. The way to do this is to 
introduce an additional shift of the one-electron potential due to the 
electrostatic interaction of the electrons inside each atomic sphere with the 
missing charge distributed outside of the sphere and postulate that the 
interaction comes from the boundary between the atomic sphere and the 
effective medium. Such a shift may be associated with an {\it intrasite\/} 
interaction, which has no connection, at all, to the effective medium.

This is exactly what is done in the locally self-consistent Green's function
method \cite{abrikosov96} where one goes beyond the single-site approximation 
for Poisson's equation by means of a supercell which models the spatial 
distribution of the atoms in a random alloy while a CPA effective medium is 
used in the electronic structure calculations beyond a local interaction 
zone (LIZ). If the LIZ consists of only one atom, the LSGF method becomes 
equivalent to the CPA method with a properly defined electrostatic potential 
and energy \cite{abrikosov96}. In this case, however, each atom in the 
supercell has its own electrostatic shift given by the Madelung potential 
from all the other atoms in the supercell while the effective medium is the 
same for all atoms. It is clear that such an additional shift for each alloy 
component does not interfere with the CPA because the CPA effective medium is 
determined on the basis of the one-electron potentials including these shifts 
and because the CPA itself does not impose any restriction on the 
one-electron potentials of the alloy components.

Following the above arguments two groups have proposed an {\it ad hoc\/} 
expression for the electrostatic shift of the one-electron potential due 
to non-zero net charges in the atomic spheres of the alloy
\cite{korzhavyi92,abrikosov92,johnson93}. Although the basic models are 
seemingly different and based on different observations, either i), the 
net charge of an impurity in a metal is screened beyond the first 
coordination shell \cite{zeller87,stefanou87} or ii), the net charge of an
alloy component is proportional to the number of the nearest neighbors of 
the opposite type \cite{magri90}, they lead to exactly the same expression 
for the one-electron potential, i.e.,

\begin{equation} \label{eq:V_SIM}
V_i = -\frac{e^2 q_i}{R_1} ,
\end{equation}
where $V_i$ is the additional electrostatic shift of the one-electron 
potential of the $i$-th alloy component of net charge $q_i$ and $R_1$ the 
radius of the first coordination shell.

In fact, the models described above are practically identical to the model  
proposed more than three decades ago \cite{treglia78,ducastelle91} to account
for charge transfer effects in the self-consistent Hartree scheme based on
the tight-binding CPA. In this scheme the variation of the $i$-th atom energy 
level, $\delta \epsilon_i$, is proportional to the corresponding charge 
transfer $q_i$, i.e., $\delta \epsilon_i = I q_i$, where $I$ is some average 
{\it intra-atomic\/} Coulomb interaction. The non-self-consistent limit 
corresponds to $I = 0$, while $I = \infty$ provides local neutrality 
\cite{ducastelle91}. In the present context one may identify 
$\delta \epsilon_i$ with $V_i$ and it therefore follows that $-e^2/R_1$ may 
be considered an intrasite Coulomb interaction.

Although there is at least some consensus concerning the definition of the 
additional electrostatic shift (\ref{eq:V_SIM}), which gives charge 
transfers quite close to the values obtained in supercell calculations 
\cite{abrikosov98}, different workers do not agree on the corresponding 
electrostatic contribution to the total energy of the random alloy. Some 
completely deny even the possibility of having such a term in a "consistent" 
SS-DFT-CPA theory \cite{gonis96,ujfalussy00} while others argue about the 
details of how this term should be defined \cite{magri90,johnson93,%
korzhavyi95,ruban95}. It would seem that the presently suggested models of
charge transfer effects in the single-site approximation to the electrostatic 
problem, except the trivial elimination of the net charges by adjusting the 
radii of the atomic spheres of the alloy components, may be considered 
neither exact nor even "a consistent theory". It is the main purpose of the
present paper to shown that a consistent SS-DFT-CPA theory including a 
correct description of the charge transfer effects does indeed exist.  

Here, we define the electrostatic shift of the one-electron potential and 
the corresponding contribution to the total energy in a form which is very 
similar to that proposed by Korzhavyi {\it et al.} 
\cite{korzhavyi92,abrikosov92} as well as Johnson and Pinski \cite{johnson93},
and which provides a practically {\it exact\/} solution to the electrostatic 
problem in the single-site approximation. The actual expression for the 
electrostatic shift in the single-site model for Poisson's equation includes 
one adjustable parameter the value of which is to be obtained in supercell 
calculations by the locally self-consistent Green's function method in which 
the Madelung problem is solved exactly. It turns out that the value of the 
adjustable parameter is practically independent of lattice structure, volume,
and alloy composition due to the fact that the screening of the electrostatic
part of the problem, in the ASA at least, is almost universal.

The paper is organized as follows. In Sec.\ II we outline the main concepts 
behind the LSGF used in this work and the details of the calculations. In 
Sec.\ III we present a pragmatic solution to the problem of finding the 
Madelung shift in SS-DFT calculations on the basis of the average values of 
the net charges and Coulomb shifts, $<q_i>$ and $<V_i>$, from supercell
calculations. We also demonstrate that the linear relation between the net 
charges, $q_i$, and the corresponding Coulomb shifts, $V_i$, of the alloy 
components discovered for metallic alloys by Faulkner {\it et al.} 
\cite{faulkner95} is practically universal in the effective medium approach 
for the Green's function. This means that the response of the electron system 
to the Coulomb field is linear and universal in such systems, and that the 
screening must be universal too. That this is indeed the case is demonstrated
in the next section where we calculate the distribution of the screening 
charge in several systems and show that it is almost independent of the 
crystal structure, the alloy constituents, and the composition. 

In Sec.\ V we present a formalism for the screened Coulomb interactions in 
the single-site mean-field approach for the electrostatic potential and 
energy and demonstrate that the conventional assumption of a vanishing 
Madelung potential and energy is not valid in general. Instead, one must 
include an additional term due to {\it intrasite\/} interactions which are, 
in fact, exactly the screened Coulomb interactions. We also discuss the 
ordering contribution to the Madelung energy and show why the screening 
contribution may be obtained in supercell calculations for ordered 
structures. The contribution from the screened Coulomb interaction to the 
generalized perturbation method and the $S^{(2)}$ formalism is also 
determined. Finally, in Sec.\ VI, we demonstrate that the total energy of 
a random alloy may be reproduced exactly in single-site CPA-DFT calculations 
with corrections due to the screening intrasite interaction.

\section{Methodology}

\subsection{Spatial ergodicity and cluster expansion}

In this paper we will consider only such alloy systems that on an underlying 
crystal lattice with perfect translational symmetry satisfy two conditions: 
i) spatial homogeneity and ii) no correlations between the one-electron 
potentials at sufficiently large distances. For the Coulomb interactions in a 
random alloy both conditions may be formulated explicitly in terms of the 
average monopole electrostatic potential $V_i$ in the atomic sphere around 
site $i$ due to the charge distribution in the whole of the remaining system.
In an ordered alloy this potential is the Madelung potential. Specifically, 
the first condition means that any average values of products of potentials 
must be translationally invariant, i.e., 

\begin{equation} \label{cond1}
<V_i V_j \ldots V_k> = <T_a (V_i V_j \ldots V_k)> ,
\end{equation}
where $T_a$ is the translation operator $T_a f(r) = f(r+a)$, and the second
condition is

\begin{equation} \label{cond2}
<V_i V_j \ldots V_k T_a(V_l V_m \ldots V_n)> = 
<V_i V_j \ldots V_k><V_l V_m \ldots V_n> ,
\end{equation}
for $a \rightarrow \infty$.

According to Lifshitz {\it et al.\/} \cite{lifshitz} the space formed by the
complete set of distinct realizations of the potential $V_i$ on the lattice,
the operator $T_a$, and the property of spatial homogeneity plays the same 
role in the theory of disorder as the phase space, the operator of dynamical 
evolution, and Liouville's theorem do in statistical mechanics. Moreover,
according to Birkhoff's ergodic theorem, for any functional $f[{V_i}]$, where 
${V_i}$ is some random realization of the potential on the lattice we have

\begin{equation}
\lim_{V \rightarrow \infty} \frac{1}{\Omega} \int_{\Omega} f[T_a{V_i}] = 
<f[{V_i}]> ,
\end{equation}
i.e., the spatial and the phase-space average are equivalent \cite{lifshitz}. 
This equation constitutes the principle of spatial ergodicity, according to
which all possible finite atomic arrangements may be realized in a single 
infinite sample if the conditions (\ref{cond1}) and (\ref{cond2}) are 
satisfied.

What makes the above principle work in practice is the fact that for 
self-averaging or "measurable" quantities, which have a well-defined limit 
when the volume of the systems approaches infinity, all the correlations
of the atomic distribution become unimportant at some distances and, hence,
the sample may be chosen finite. This may be formulated explicitly by means
of the cluster expansion theorem \cite{sanchez84} which defines the 
corresponding measurable quantity in terms of the site occupation
correlation functions

\begin{equation} \label{clexp}
\Pi = \pi_0 + \sum_f \pi_f \xi_f ,
\end{equation}
where $\pi_q$ are the coefficients or interaction parameters,
$\xi_f = <\delta c_i \delta_j \ldots \delta c_k>$ the correlation function
of the figure or cluster $f$ which corresponds to a specific position of the
sites $i$, $j$, and $k$ in the lattice, and $\delta c_i$ = $c_i - <c_i>$
is the fluctuation of the site occupation numbers $c_i$ taking on values 0 
and 1 depending on whether site $i$ is occupied by one or the other 
component.

According to (\ref{clexp}) there are two practical ways of calculating the 
properties of a random system for which we have $\xi_f=0$ and, thus, 
$\Pi_{rand} = \pi_0$: i) the cluster or supercell approach, where $\xi_f=0$ 
is satisfied on average only for those clusters $f$ for which 
$\pi_f \neq 0$, or ii) the effective medium approach, which directly gives
$\pi_0$ from some knowledge of the alloy components. The first approach
is realized, for instance, in the so-called special quasirandom structure
(SQS) method \cite{zunger90} while the second approach is realized by the 
coherent potential approximation where the real atoms are substituted by a
specifically chosen effective medium on the lattice.

\subsection{The LSGF method: a combined supercell-effective medium approach}

The supercell and the effective medium approaches are combined into a single
computational scheme in the locally self-consistent Green's function (LSGF) 
method \cite{abrikosov96,abrikosov97}. In the LSGF method the supercell 
approach is used to provide the correct solution to the Madelung problem for 
a given alloy modeled by an appropriate supercell. It is also used in part 
in the electronic structure calculations, which are performed separately for 
every atom in the supercell by means of the local interaction zone centered 
at each atom. Inside the LIZ the multipole scattering equations are solved 
exactly, while the region outside the LIZ is represented by the effective 
medium, which is usually taken to be the CPA effective medium built on all 
the one-electron potentials in the supercell. This means that every atom of 
the supercell "sees" only the CPA effective medium outside the LIZ, which  
according to the CPA definition represents a random alloy. In other words, 
the one-electron Green's function of the supercell (sc) obtained in LSGF 
calculations may be presented as \cite{abrikosov96,abrikosov97}

\begin{equation} \label{G_LSGF}
G^{sc} = \sum_i [G^0_i + \sum_f \Delta G_{if} \xi^{LIZ}_{if} ] ,
\end{equation}
where $G^0_i$ is the Green' function of the $i$th atom in the supercell
embedded in the CPA effective medium, $\Delta G_{if}$ the contribution to
$G^0_{i}$ due to the presence of the specific atomic arrangements on the 
figure $f$ in the LIZ as specified by the correlation function 
$\xi^{LIZ}_{if}$, which is equal to that of the supercell $\xi_{if}^{sc}$ 
if the figure $f$ is circumscribed by the LIZ, i.e., it can be put inside 
the LIZ in such a way that one of its vertices coincides with the central 
atom of the LIZ, otherwise $\xi^{LIZ}_{if}$=0.

Equation (\ref{G_LSGF}) clearly shows how the LSGF works, and in particular,
how the effective medium approach, represented by $G^0_i$, is combined with
the cluster or supercell approach, represented by by the second term. It 
follows from (\ref{G_LSGF}) that the LIZ allows one effectively to cut off 
the contributions from the clusters which are not circumscribed by the LIZ 
(an equivalent formulation in terms of effective interactions is given in 
Ref.\ \cite{abrikosov97}). If the LIZ is single-site, i.e., it consists of 
only one atom (LIZ=1), the contributions from the second term in 
(\ref{G_LSGF}) vanish and we are left with the usual CPA or pure effective 
medium approach to the electronic structure problem. It is this single-site
approach, referred to as SS-LSGF, which will be used in most of the present 
work. It has the advantage over the usual SS-DFT-CPA method that Poisson's 
equation is solved exactly within a given approximation for the form of the 
electron density. 

Note, however, that the LSGF method by no means is restricted to the 
single-site approximation. In fact, it allows us to include local 
environment effects in the electronic structure calculations for the figures 
circumscribed by the LIZ if on average $\xi^{<LIZ>}_f = \xi_f^{sc} = 0$ 
which is the case in a random alloy. In this respect the LSGF method may be 
considered a {\it self-consistent\/} embedded cluster method (ECM) of the 
kind proposed by Gonis {\it et al.} \cite{gonis77} more than two decades ago.
With a proper choice of the supercell used to model a given random alloy 
\cite{abrikosov97} the LSGF solves two major problems of the ECM: i) it 
provides a set of clusters to represent an alloy with a given short range 
order and ii) it allows one to close the DFT loop with the correct treatment 
of the electrostatics.

There is one important point concerning the electronic structure obtained
in the LSGF with the CPA effective medium which should be mentioned: Despite
the fact that it can be quantitatively accurate, it is {\it qualitatively\/}
different from the electronic structure which would result from direct 
supercell calculations with periodic boundary conditions. That is, the 
electronic spectrum in the LSGF-CPA method is always complex, unless all the 
atoms are equivalent in the supercell (pure metal) or the size of the LIZ is 
infinite. Thus, in the LSGF-CPA the electronic structure of an {\it ordered\/} 
alloy is never correct, although it may be calculated with arbitrarily high 
accuracy. On the other hand, since Blochs' theorem is not applied to the
supercell during the electronic structure calculations, the LSGF method is
a perfect tool for calculating SQS as opposed to the ordinary band structure
methods, which in this case lead to a real, i.e., qualitatively incorrect,
electronic structure of the random alloys due to the fact they present purely
a supercell approach.

\subsection{The choice of the supercell in the LSGF and details of the
calculations}

It is possible to obtain the Madelung potential and energy by a combined
supercell (cluster) and effective medium approach similar to that used in 
the Green's function approach to the electronic structure problem within 
the LSGF method. However, this requires some knowledge of the charge-charge 
correlations or the screening in the alloy. Hence, the only way to solve the 
problem is to use a supercell model with the Madelung potential and energy 
determined exactly from the {\it bare\/} electrostatic interactions, as it 
is usually done. Here, another problem arises: The supercell should be
constructed such as to provide zero correlation functions up to the distance 
where the net charges of the alloy components become uncorrelated or 
completely screened.

In the calculations presented below we assume at the outset the existence of 
a short-range screening which occurs over the distance of the first several 
coordination shells. This assumption is based on results obtained by the 
charge-correlated model \cite{magri90}, on single-impurity calculations 
\cite{zeller87,stefanou87} as well as on the most recent LSGF calculations 
by Ujfalussy {\it et al.\/} \cite{ujfalussy00}. The latter authors 
demonstrated that a 16-atom supercell for an fcc equiatomic random alloy, in 
which the SRO parameter at the 8'th coordination shell must be equal to 1 
due to the translation symmetry, i.e., all the atoms in the 8'th coordination
shell are the same as that at the central site, yields practically the same 
average charge transfer and total energy as a 250-atom supercell, in which 
the SRO parameter at the 8'th coordination shell corresponding to a random 
number generator distribution of the alloy components on the lattice should 
be $<<$ 1 (see also \cite{wolverton96}).

Thus, in all the random alloys considered below the distribution of the atoms
in the supercell was chosen such that the SRO parameters (or pair correlation
functions, $\xi^{(2)}_f$) were exactly zero at least in the first 6 
coordination shells and small (not greater 0.01 in absolute value) up to the 
8th coordination shell. Although the multisite correlation functions have not
been optimized, they should not play a significant role in the ASA where only
monopole intersite Coulomb interactions are taken into account. 

The electronic structure calculations were in all cases performed in the 
scalar relativistic approximation by the KKR-ASA technique \cite{ruban99}
with an $s$-, $p$-, and $d-$basis in the framework of either the usual
SS-DFT-CPA or the LSGF methods with a CPA effective medium. The ASA
(no multipole corrections to the Madelung potential and energy) has been used
in the electrostatic part of the problem. The integration of the Green's 
function over energy was performed in the complex plane over 16-20 energy 
points on a semicircular contour. The local density approximation was used in
the DFT part with the Perdew and Zunger \cite{perdew81} parameterization of 
the results by Ceperly and Alder \cite{ceperly80}.

\section{Net charge and Madelung potential in metallic alloys}

Here we discuss a pragmatic solution to the following problem: Can one 
devise a Madelung potential for the alloy components to be used in 
SS-DFT-CPA calculations such that the charge transfer effects, i.e., 
the net charges of the alloy components, are consistent with those obtained 
in SS-LSGF calculations where charge transfer effects are treated properly?
The fact, that such a potential can be found, may seem surprising in view of
the principle difference between the LSGF and the SS-DFT-CPA methods. In the
LSGF approach all the atoms in the supercell are different due to their 
different local environment while in the usual SS-DFT-CPA approach one deals
only with average quantities, i.e., in terminology of the Ref.\ 
\cite{faulkner98}, the LSGF supercell approach is equivalent to the 
{\it polymorphous} model of the alloy while the effective-medium approach
is equivalent to the {\it isomorphous} model. However, it is obvious, that
this can be done {\it on average}.

\begin{figure}
\centerline{\psfig{figure=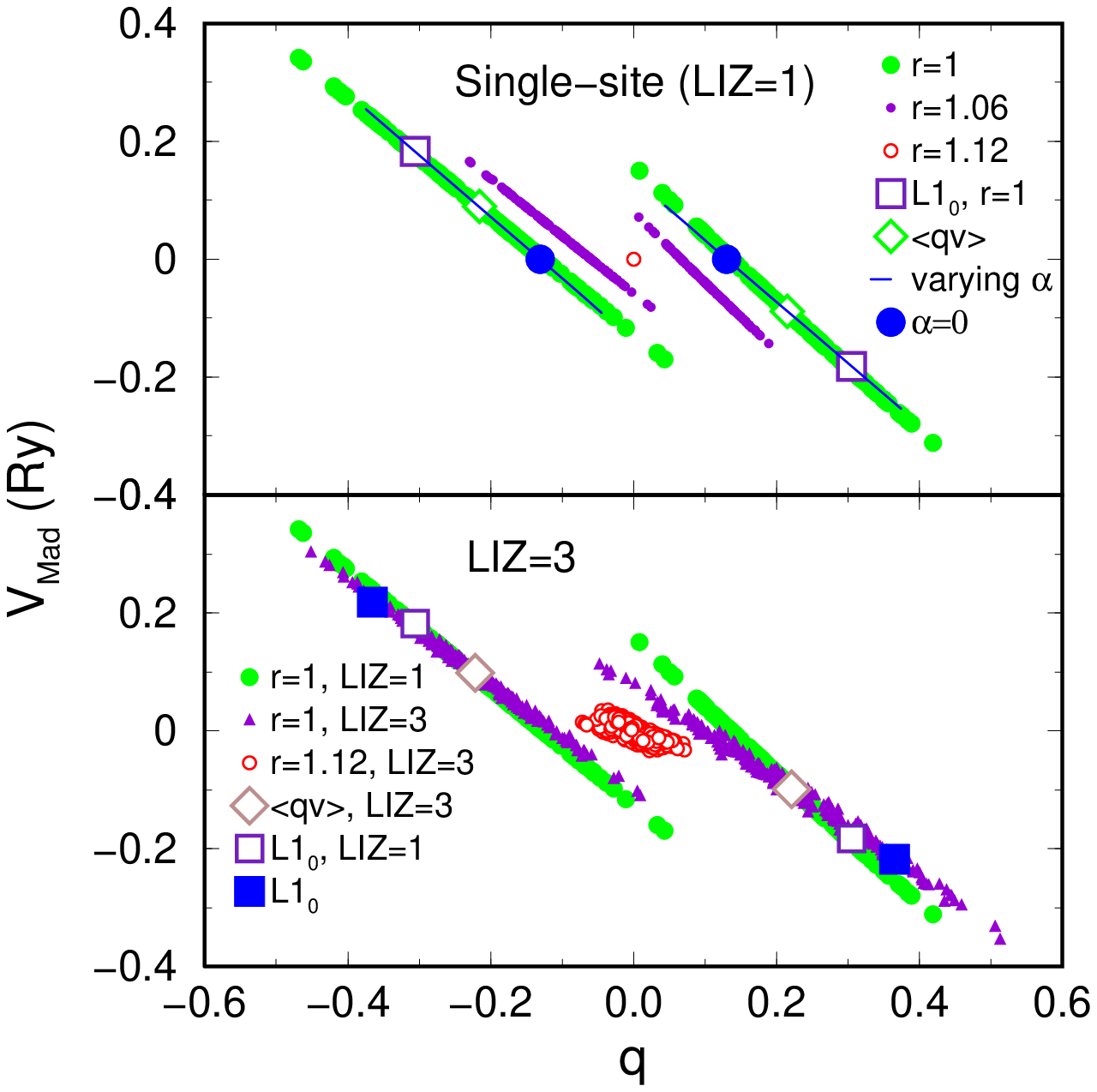,height=10.0cm}}
\caption[1]{The distribution of the net charges, $q_i$, and corresponding
Madelung potentials, $V_i$ in the 512-atom supercell, modeling a random
Al$_{50}$Li$_{50}$ alloy, ordered L1$_0$ alloy and in the single-site
CPA-DFT calculations obtained by varying $\alpha$.}
\label{qV_AlLi}
\end{figure}

It was discovered by Faulkner {\it et al.} \cite{faulkner95} from supercell 
calculations that the net charges on different sites $i$, $q_i$, and the 
corresponding Madelung potentials, $V_i$, obey a linear relationship. In 
Fig.\ \ref{qV_AlLi} we show such a $qV$ relation for a 512-atom supercell
which models a random Al$_{50}$Li$_{50}$ alloy on an underlying fcc lattice.
For comparison we also show the charge and Madelung shift for AlLi in the
ordered L1$_0$ structure. All results are obtained by LSGF calculations with 
the CPA effective medium in i) the single-site approximation for the 
electronic part of the problem, SS-LSGF, i.e., LIZ=1, (upper panel) and ii)
with the perturbation in the electronic structure caused by the local 
environment up to the second coordination shell, i.e., embedded 
cluster(EC)-LSGF, LIZ=3,(lower panel), included in the Green's function. 
For random alloys the inclusion of more distant coordination shells do not 
affect the results significantly and thus the LIZ=3 results may be 
considered to be converged in the LIZ size.  

The most striking feature of the $qV$ relation obtained in the SS-LSGF 
calculations is the perfect alignment of the $qV$ points along two almost 
straight lines, one for each alloy component. This is, in fact, very similar 
to what has been observed by Pinski \cite{pinski98} in model calculations 
using the Thomas-Fermi approximation. Furthermore, a change of the ratio of 
the atomic sphere radii of the alloy components, $r=S_{Al}/S_{Li}$, leads to 
a rescaling of the $qV$ points. Hence, for a specific ratio, $r=1.12$ in the 
present case, the $qV$ relation collapses into the single point: 
$(q,V)=(0,0)$. The existence of this point in the SS-LSGF is a consequence 
of the fact that all Al atoms as well as all Li atoms become 
indistinguishable if the net charges of the alloy components are zero:
The difference between the atoms caused by local environment effects is 
solely due to the Madelung shift, which is zero in this case. Thus, for this 
particular choice of $r$ the polymorphous model is identical to the usual 
isomorphous model, or the SS-LSGF method is identical to the SS-DFT-CPA
method.

On the other hand, it is clear that the two models are not equivalent when 
local environment effects are included in the electronic structure part
of the LSGF method, i.e, for LIZ$>$1. This is demonstrated in the lower panel 
of Fig.\ \ref{qV_AlLi} where the local environment effects are clearly seen 
to destroy the strict alignment of the $qV$ points and, as consequence, the 
possibility of choosing electroneutral atomic spheres by a single $r$ value. 
However, even if this were possible, all the atoms, or the corresponding
one-electron potentials, would still be different.

The discussion of local environment effects is beyond the scope of the 
present paper, and the results are included only to demonstrate the {\it 
qualitative\/} difference between the correct results and those obtained by 
the LSGF method in the single-site approximation: Inclusion of intersite 
correlations in the electronic structure calculations leads to a real 
polymorphous description of random alloys which cannot be mimicked by a 
single-site LIZ. As a consequence, as we will show later the SS-DFT-CPA
method can reproduce the results of the SS-LSGF exactly, but will, in 
general, reproduce only approximately the correct solution to the supercell
or polymorphous model of a random alloy.

The two $qV$ points for the ordered L1$_0$ structure are seen to fall on the 
$qV$ lines for the random alloy as already noted in Ref.\ \cite{faulkner95}),
and, in fact, all the points on the $qV$ relation obtained by the SS-LSGF 
method may be reproduced by a series of ordinary SS-DFT-CPA calculations, 
by using the shift of the one-electron potential defined in a way similar to
(\ref{eq:V_SIM}), i.e., 

\begin{equation} \label{eq:V}
V_i = -\alpha \frac{e^2 q_i}{S} ,
\end{equation}
where $q_i$ is the net charge of the alloy components, $S$ the Wigner-Seitz 
radius, and $\alpha$ a parameter which may be varied arbitrarily in the 
SS-DFT-CPA calculations without specifying its physical meaning. However, it 
is important to note, that $\alpha=-\infty$ corresponds to the electroneutral 
case ($q_i=0$) and $\alpha=0$ to the limit where there is no response of the 
system to charge transfer effects. As we will see later, the values of the 
net charges $q_{0i}$ obtained in the SS-DFT-CPA calculations with
$\alpha=0$ are important scaling parameters. It is also useful to note, that 
for the L1$_0$ structure $\alpha_{L1_0}$=0.8811575 \cite{landa91}, and in 
the screened impurity model (\ref{eq:V_SIM}) $\alpha_{SIM} =$ 0.552669 and 
0.568542 for the fcc and bcc crystal structures, respectively.

Fig.\ \ref{qV_AlLi} shows the $qV$ relation, indicated by the black line,
obtained in the SS-DFT-CPA calculations including (\ref{eq:V}) with $\alpha$
varying from $-$1.5 to 5 together with the SS-LSGF results, gray circles. 
It is clearly seen that the {\it "isomorphous"} and {\it polymorphous} $qV$
relations coincide, and this allows one to make an isomorphous model 
consistent with the polymorphous results. The point is that all the net 
charges and corresponding Coulomb shifts in the polymorphous model have 
significance only in terms of the average values they produce. This is so, 
because every supercell has its own set of net charges and Madelung shifts 
and, in the case of an infinite system, there is an infinite number of 
different $qV$ points. Their average values, $<q_i>$ and $<V_i>$, however, 
have a well-defined physical meaning as conditional averages of  
self-averaging quantities, and thus it is the average $<qV>$ point which must 
be reproduced by the isomorphous model. Hence, for a random alloy $\alpha$ 
is given by

\begin{equation} \label{eq:alpha}
\alpha_{rand} = -\frac{S}{e^2} \frac{<V_i>}{<q_i>}.
\end{equation}
Note, that in a binary AB alloy, it clearly does not matter, for which alloy 
component, $i=A,B$, $\alpha_{rand}$ is determined, since 
$<V_{A}>/<q_{A}> = <V_{B}>/<q_{B}>$ \cite{faulkner95}. The same is true for
multicomponent alloys, but in this case, rather than being a trivial 
consequence of the charge neutrality condition, it follows from the physical 
origin of $\alpha_{rand}$, which will be discussed in the next two sections.
For an fcc Al$_{50}$Li$_{50}$ alloy at $S$=2.954 a.u. we find from the LSGF 
calculations $\alpha_{rand}=$0.60716. 

\begin{figure}
\centerline{\psfig{figure=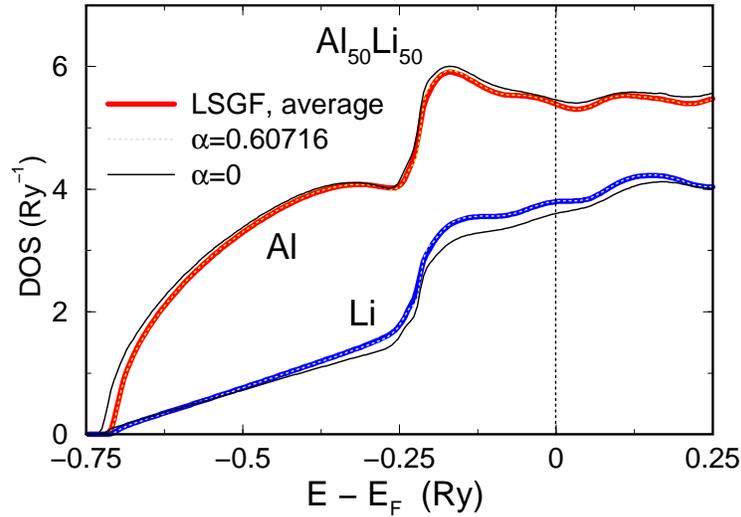,height=7.0cm}}
\vspace{1cm}
\caption[2]{The local density of states in the Al$_{50}$Li$_{50}$ obtained by the
LSGF method with LIZ=1 and by the single-site CPA-DFT with different values
of $\alpha_{rand}$.}
\label{DOS}
\end{figure}

Of course, the coincidence of ($<q_i>,<V_i>$) is a necessary, but not 
sufficient condition for the equivalence of the isomorphous and polymorphous 
models. The two models may be called equivalent only if the electronic 
structure of the random alloy and its conditional averages agree. In Fig.\ 
\ref{DOS} we show that this is indeed the case: The local densities of 
states (DOS) for the Al and Li atoms in Al$_{50}$Li$_{50}$ calculated by 
the SS-DFT-CPA method coincide with the corresponding conditional
average state densities obtained in the SS-LSGF calculations for 
the 512-atom supercell. For comparison we also show the DOS obtained with 
$\alpha=$0 corresponding to the "conventional" CPA. Although the latter 
differs from the correct state density, it is obvious, that the neglect of 
the electrostatic shift (\ref{eq:V}) has only a minor effect on the DOS.

\begin{figure}
\centerline{\psfig{figure=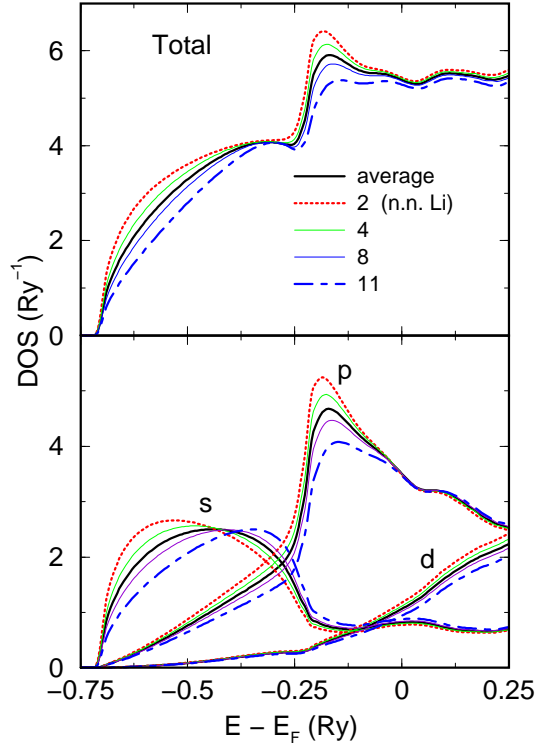,height=10.0cm}}
\caption[3]{The site- and $l$ projected density of states for Al atoms 
in Al$_{50}$Li$_{50}$ having different Madeulng potentials due to different
numbers of nearest neighor Li atoms.}
\label{DOS2}
\end{figure}

The reason why the average state densities coincide is the following: In the
SS-LSGF method the difference between the atoms of the same type comes only 
through the corresponding Madelung shift. A shift in potential leads to a 
change in the charge transfer through a {\it skewing \/} of the local DOS 
as seen in fig.\ \ref{DOS2}. Therefore, when the conditionally averaged DOS 
is obtained, the skewing contributions from the individual atomic sites 
caused by $V_i$ will cancel and leave only the DOS given by the average 
$<V_i>$. Of course, this is true only in the SS-LSGF method (LIZ=1). In 
fact, the local environment effects in concentrated random alloys may 
influence quite strongly the electronic structure of the central site of 
the LIZ.

\begin{figure}
\centerline{\psfig{figure=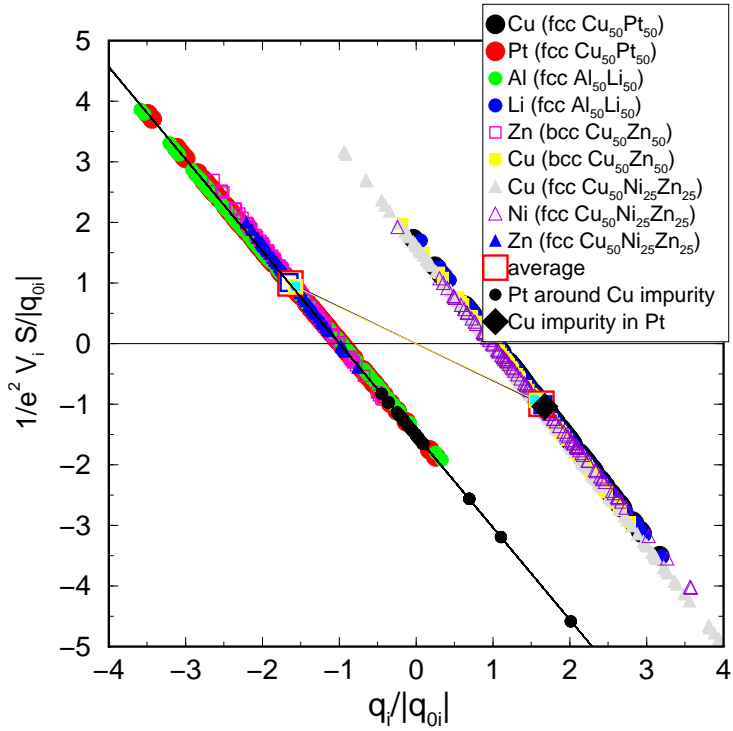,height=10.0cm}}
\caption[4]{$qV$ relation scaled by $q_{0i}$, and the Wigner-Seits radius, $S$.}
\label{qV_univ}
\end{figure}

To investigate how the $qV$ relation depends on the system we show in Fig.\ 
\ref{qV_univ} $qV$ relations for five different systems including a Cu 
impurity in Pt ($S=$ 3 a.u.) and four random alloys: 
fcc Cu$_{50}$Pt$_{50}$ ($S=$ 3 a.u.), 
fcc Al$_{50}$Li$_{50}$ ($S=$ 2.954 a.u.), 
bcc Cu$_{50}$Zn$_{50}$ ($S=$ 2.7 a.u.), and
ternary fcc Cu$_{50}$Ni$_{25}$Zn$_{50}$ ($S=$ 2.65 a.u.). In the plot all 
charges have been normalized by $q_{0i}$ obtained in the no response limit, 
i.e., $\alpha=0$ or $V_i= 0$, and all Madelung shifts have been normalized 
by $q_{0i}/S$. To partly simplify the plot we have used $|q_{0i}|$ in the 
normalizations, thereby separating the $qV$ relations into two lines rather 
than one.

The results presented in Fig.\ \ref{qV_univ} show the existence of a 
universal $qV$ relation. Or in other words, the linear-response function
$\chi$ which gives the change in the net charge relative to $q_{0i}$ caused
by $V_i$, i.e., $q_i - q_{0i} = 1/e^2 \chi V_iS$, is a universal constant in 
metallic alloys in the ASA. From the results presented in the figure we find
that $\chi \approx -$0.63. This unavoidably leads to the existence of a 
single, unique $\alpha_{rand}$ as witnessed by the coincidence of all the 
$<qV>$ points in Fig.\ \ref{qV_univ}. Strictly speaking, the slopes of the 
$qV$ lines are not exactly identical and, in fact, $\alpha_{rand}$ varies 
from 0.6 in Cu$_{50}$Zn$_{50}$ and Cu$_{50}$Ni$_{25}$Zn$_{50}$ to 0.615 in 
LiMg alloys, not included in the figure. However, for most practical purposes
the choice $\alpha_{rand}= 0.607$ provides a sufficiently accurate 
description of the electronic structure of random alloys in the SS-DFT-CPA 
method in the ASA for the electrostatic part.

\section{Screening charge in metallic alloys}

In the previous section we have, in effect, defined a procedure whereby  
SS-DFT-CPA calculations may provide the exact solution to the
electrostatic problem in random alloys. The only requirements are that the 
Madelung shift (\ref{eq:V}) is included and that the constant $\alpha_{rand}$ 
is obtained from (\ref{eq:alpha}) with the average Madelung potential and net 
charges of the alloy components determined in supercell calculations by the 
SS-LSGF method. In the derivation of the procedure we have used some general 
arguments which do not clarify the physical origin of the universal value of 
$\alpha_{rand}$. However, it is clear, that (\ref{eq:V}) accounts for the 
missing charge $-q_i$ in the single-site Poisson equation for the $i$'th 
atomic sphere. Thus, $\alpha_{rand}$ must be connected to the screening. 

The linear character of the $qV$ relation indicates that the screening in the
impurity case as well as in the case of a random alloy may be very well
described by linear response theory. Owing to enhanced electron scattering 
at opposite regions of the Fermi surface linear response predicts in the case 
of a free-electron gas the existence of long-range Friedel oscillations,
which however decrease relatively fast ($\sim r^{-3}$) with the distance. In 
a random alloy, on the other hand, the screening is much more efficient due 
to the finite life time of the Bloch states for the underlying crystal 
lattice and the spatial distribution of the screening density decays 
exponentially. In this respect charge correlated model 
\cite{magri90,wolverton96} adopted by Johnson and Pinski \cite{johnson93} in 
the cc- and scr-CPA method or the equivalent screened impurity model 
\cite{abrikosov92,korzhavyi95,ruban95} may be viewed as the first 
approximations for the screening. 

Based on the fact, that a single impurity in a metallic host is a particular
case of a dilute random alloy, one would expect, and the results for a single 
Cu impurity in Pt presented in the previous section unambiguously 
indicate this, that the screening effects in the two cases are similar. It
is therefore surprising that Faulkner {\it et al.} \cite{faulkner98} and 
Ujfalussy {\it et al.} \cite{ujfalussy00} claim that the screening in a 
random alloy is {\it qualitatively}\/ different from that found in a single 
impurity system. In fact, these authors found extremely long-range 
correlations between the Madelung potential at some particular site and the 
net charges at the other sites. Since the Madelung shift on a site is 
proportional to the net charge on the site, this may happen only if there are 
extremely long-ranged correlations between net charges or, in other words, 
there is no screening. However, this result has been obtained on the basis of
summations of the direct or {\it bare}\/ Coulomb interactions which is, at 
best, an ill-defined procedure, even mathematically. 

To clarify the issue of screening we will perform the following computer 
experiment which will allow us to establish the range of the net-charge 
correlations or the screening in random alloys for one particular site. We 
set up a 512-atom supercell which represents an fcc Cu$_{50}$Pt$_{50}$ random 
alloy (all SRO parameters are equal to zero up to the sixth coordination 
shell and $\sim$ 0 for at least the next 10 coordination shells) and perform 
self-consistent SS-LSGF (LIZ=1) and EC-LSGF (LIZ=3) calculations. We then
substitute one Pt atom with one Cu atom in some site which, in general, may 
be chosen arbitrarily. However, to keep the atomic distribution as close as 
possible to the random distribution we chose a site the local environment of 
which corresponds to the random alloy (having equal number of Cu and Pt 
atoms) for the first three coordination shells. We then repeat the 
self-consistent LSGF calculations for the supercell with the substituted atom 
and find new values for the net charges in the supercell. It is clear that, 
when LIZ=1, the difference between the net charges in the two calculations, 
$\Delta q_i$, gives the charges induced by the change of the net charge at 
the substitution site (in the case of LIZ=3 the local environment effects 
also effect the charge transfer). This charge is simply the screening 
charge.

\begin{figure}
\centerline{\psfig{figure=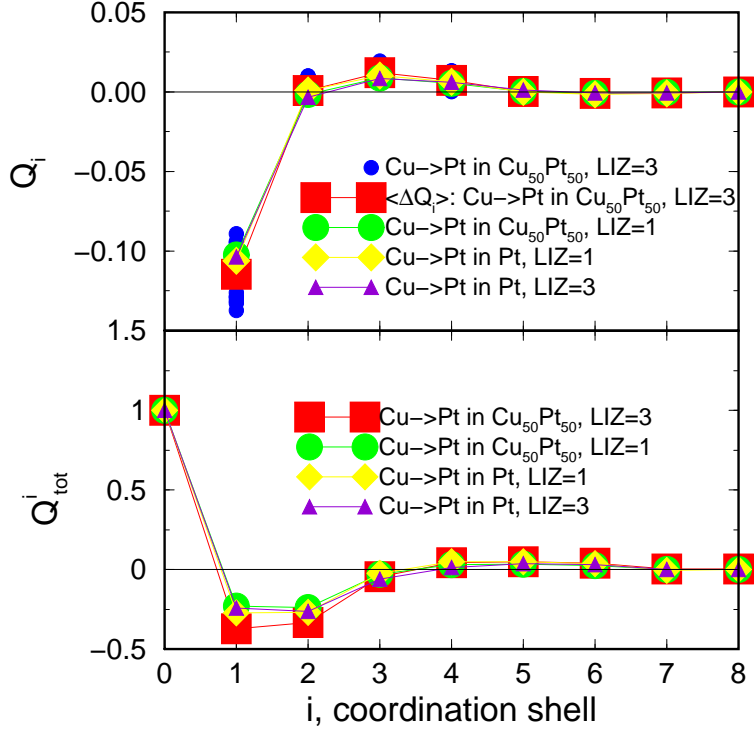,height=10.0cm}}
\caption[5]{The distribution of the screening charge in the random
Cu$_{50}$Pt$_{50}$ alloy and in Pt due to substitution of a Cu atom.}
\label{q_scr}
\end{figure}

In the upper panel of Fig.\ \ref{q_scr} we have plotted the normalized, 
induced charges 

\begin{equation} \label{Q_i}
Q_i = \frac{\Delta q_i}{\Delta q_0}
\end{equation}
at the first eight coordination shells around the substitution site $i=0$ for
Cu-Pt substitution in a Cu$_{50}$Pt$_{50}$ random alloy and for Cu-Pt 
substitution in pure Pt ($S=$ 3 a.u.) obtained in the single-site 
approximation for the electronic structure (LIZ=1) as well as with local 
environment effects included (LIZ=3). One may see that, while the local 
environment in a pure metal hardly affects the distribution of the net 
charges, it does introduce a dispersion in the distribution of the net 
charges in the random alloy, which is quite substantial at the first
coordination shell but which practically disappears beyond the 5'th 
coordination shell where, in fact, all the induced charges almost vanish. 

To demonstrate that the net charge of the Cu "impurity" indeed becomes 
screened we show in the lower panel of Fig.\ \ref{q_scr} the total normalized 
induced charge in the $i$'th shell

\begin{equation}
Q^i_{tot} = \sum_{j=0}^i z_j Q_j ,
\end{equation}
where $z_j$ is the coordination number of the $j$'th shell. I is seen that
$Q_i$ vanishes beyond the 7'th coordination shell in all cases, and we 
conclude that the screening in a random alloy in the single-site 
approximation is practically the same as the screening in the case of a 
single impurity in a pure metal. There is neither qualitative nor 
quantitative differences between the impurity and the alloy cases.

\begin{figure}
\centerline{\psfig{figure=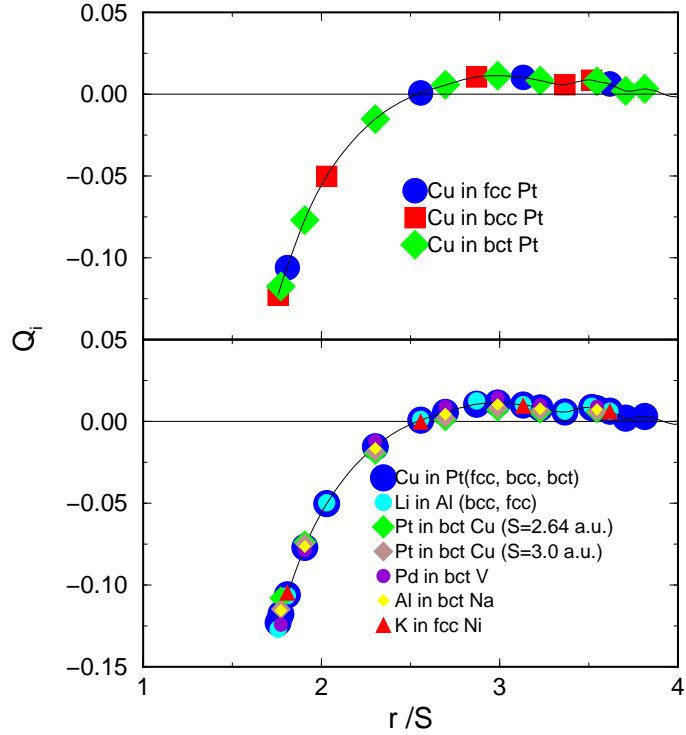,height=10.0cm}}
\caption[6]{The distribution of the screening charge in different metals
having different crystal structure and lattice parameter.}
\label{q_scr_all}
\end{figure}

In the upper panel of Fig.\ \ref{q_scr_all} we show the distribution of the
screening charge (not to be confused with the screening {\it density}: the 
screening charge is, in fact, the screening density integrated in the
corresponding atomic sphere) for a Cu impurity in fcc, bcc, and bct Pt 
plotted as a function of the distance from the impurity site in units of the 
Wigner-Seits radius, $S$. It is clear that the screening charge follows a 
single, common curve which does not depend on the structure. In fact, by 
changing the $c/a$ ratio in the bct structure one may completely fill the 
remaining gaps in the calculated curve. In the lower panel of the figure we 
have collected the results for the distribution of the screening charges in 
seven different systems including such hosts as Pt, Al, Cu, V, Na, and K. It 
appears that the screening in metallic alloys depends neither on the crystal 
structure nor on the nature of the alloy components, at least, when described 
within the ASA. 

The universal picture of the screening in alloy systems found above is partly
destroyed when the electrostatics is treated more correctly, for instance by 
including multipole moment contributions to the one electron Madelung 
potential and energy. However, the ASA still gives a qualitatively correct 
picture and catches the main physics behind the phenomenon. Hence, it is 
worth to discuss the origin of such a universality in both the screening and 
the response function. 

First of all, it was understood long time ago that the net charges in the 
atomic spheres of the alloy components has very little in common with the 
"charger transfer" in terms of the redistribution of the electron charge 
between the alloy components (see, for instance Ref.\ \cite{watson91}). Even 
in the case of the so-called ionic solids the self-consistent charge 
distribution is very close to that obtained from a linear superposition of 
the free-atom electron-densities \cite{averill90} and this is the reason for 
the success of the charge-correlated model \cite{magri90}, in which the net 
charge is proportional to the number of nearest neighbors of the opposite 
type (see also Ref.\ \cite{wolverton96}).

What we are seeing is basically a size effect: The net charges originate
from the redistribution of the electron density in the {\it interstitial} 
region between the atomic spheres. The electron density in metals and their 
alloys in this region is very smooth and may be well described by a
free-electron model, even for transition metal alloys. The interstitial 
density is much easier to perturb than the density closer to the atomic 
nuclei and it participates in the screening. On this basis one may, in fact, 
developed a model based on linear response theory which leads to a 
semi-analytical description of the universal screening. However, this is
beyond the scope of the present paper.

The calculated distribution of the screening net charge may be used to obtain 
the screening contribution to the one-electron potential in the single-site 
model for the Poisson equation given by (\ref{eq:V}) with $\alpha$ equal to

\begin{equation} \label{eq:alpha_scr}
\alpha_{scr} = \frac{S}{e^2}\sum_i z_i \frac{Q_i}{R_i} ,
\end{equation}
where $R_i$ is the radius of the $i$'th shell with coordination number $z_i$. 
Using the results for the screening charge, $Q_i$, in the case of Cu-Pt 
substitution for one particular site (not on average!) in the 
Cu$_{50}$Pt$_{50}$ random alloy we find from (\ref{eq:alpha_scr}) after 
summation up to the 8'th coordination shell that $\alpha_{scr}$ = 0.60572. 
At the same time, the average values of the net charges and the Madelung 
potentials, $<q>$ and $<V>$, in conjunction with (\ref{eq:alpha}) gives 
$\alpha_{rand}$ = 0.60530. That is, $\alpha_{rand} = \alpha_{scr}$. Since, 
there is only one effective medium in the supercell LSGF calculations, it is 
obvious that the screening is the same for all the sites and, thus, should 
not depend on the alloy component, that is, the ratio $<V_i>/<q_i>$ does not 
depend on the alloy components.

It is important to understand that $\alpha_{rand}$ appears in the formalism 
due to the {\it intrasite}\/ interactions between the electron density inside
an atomic sphere and its screening (or missing) charge. Therefore the 
inclusion of (\ref{eq:V}) in SS-DFT-CPA calculations does not {\it in any 
way}\/ contradict the "conventional" mean-field picture according to which 
the  contribution from {\it intersite}\/ interactions to the Madelung 
potential is zero in the SS-DFT-CPA. In fact, neglecting the term will lead 
to incorrect results. On the other hand, the parameter $\alpha_{L1_0}$ 
which allows one to perform SS-DFT-CPA calculations {\it identical}\/ to 
those of the SS-DFT-CPA for the L1$_0$ structure, has an origin different 
from that of $\alpha_{scr}$ determined by (\ref{eq:alpha_scr}). The former is 
a constant, which apart from the $L1_0$ symmetry does not depend on anything,
and in particular, not on the screening, while the latter is a system 
dependent parameter which is entirely determined by the type of intersite 
interactions or the screening in a given system.

This means that there is no connection between the Madelung constants for 
ordered structures and $\alpha_{scr}$. The reason why it is possible to 
obtain $\alpha_{scr}$ from the supercell calculations for the completely 
ordered structures, as we have just done, is the fact that at the large  
distances, where the atomic-distribution correlation-functions are not zero 
anymore, the "real" {\it intersite} interactions do not contribute to the 
electrostatics due to the short-ranged screening. We will return to this 
point in the discussion of the Coulomb energy of a random alloy, but here we 
would like to comment on the use of the single-site approximation in the
Green's function calculations.

Our LSGF calculations of impurities in different metals indicate that the 
problems observed in the single-site Green's function impurity calculations
by Stefanou \cite{stefanou87} and Drittler {\it et al.} \cite{drittler89}
do not originate from the single-site approximation for the Dyson equation,
but from Poisson's equation, which these authors also solve in the
single-site approximation. The use of the screening electrostatic shift for 
the one-electron potential allows one to solve the impurity problem in the
single-site Green's function formalism in the ASA or in the MT approximation
almost exactly. This is so because the impurity case corresponds to the 
dilute limit of a random alloy where the concentration of one of the alloy 
components approaches zero. In this case, the contribution to the electronic 
structure due to the local environment effects becomes negligible, and the 
electronic structure of the impurity obtained by the single-site or the
cluster Dyson equations become almost identical. Such an effect may be seen, 
for instance, in Fig.\ \ref{q_scr}: The dispersion of the screening net 
charge found in the concentrated alloy case, i.e., Cu$_{50}$Pt$_{50}$ 
(LIZ=3), vanishes in the case of a Cu impurity in Pt. The effect is, in 
fact, the origin of the increasing accuracy of the CPA with decreasing 
concentration of one of the alloy components.

\section{Screened Coulomb interactions}

\subsection{The Madelung energy of a random alloy in the single-site 
mean-field approximation}

The formalism describing the electrostatics of random alloys in the 
single-site mean-field approximation, where all the A and all the B atoms
are represented only by the appropriate conditional averages, is indeed a 
trivial one, although it has remained quite a confusing issue for more than 
30 years, starting from the definition of the electrostatic energy of a 
random alloy given by Harrison in 1966 in connection with pseudopotential 
theory \cite{harrison66}. According to Harrison the Madelung energy, which 
may be associated with the electrostatic interactions of the net charges of 
the ions of a random A$_c$B$_{1-c}$ alloy, is given by \cite{harrison66}

\begin{equation} \label{eq:E_Harr}
E_{Mad}^{rand-ss} = -\frac{e^2}{2} \frac{\alpha_{0}}{S} 
c(1-c)(Z_A - Z_B)^2 +
\frac{e^2}{2} \frac{\alpha_M}{S} \tilde{Z}^2 ,
\end{equation}
where $Z_i$ are the ion charges of the alloy components, $\alpha_M$ is the 
Madelung constant of the underlying lattice, $S$ the radius of the 
Wigner-Seitz sphere, $\tilde{Z}$ the average charge equal to 
$cZ_A + (1-c)Z_B$, and $\alpha_0$ some constant.

Equation (\ref{eq:E_Harr}) was, in fact, not derived specifically for Coulomb 
interactions. Rather, it was suggested on the basis of a more general 
consideration of the band-structure contribution to the total energy of a 
random alloy within second order perturbation theory. Later, it was shown
by Krasko \cite{krasko70} that in a system with randomly distributed A and B 
ions, having charges $Z_A$ and $Z_B$, respectively, embedded in a medium of 
compensating charges, the electrostatic energy included only the second term
in (\ref{eq:E_Harr}) while the first term vanished.

It is obvious that (\ref{eq:E_Harr}) is valid also in the CPA-DFT if one
substitutes the ion charges $Z_i$ by the net charges $q_i$ of the atomic 
spheres. In the ASA the average charge $cq_A + (1-c)q_B$ is zero, and in this 
particular approximation the contribution to the Madelung energy of a random 
alloy given by the second term in (\ref{eq:E_Harr}) will also vanish. Note, 
however, that the term vanishes neither in inhomogeneous systems 
\cite{ruban95,ruban99} nor in the MT-approximation \cite{johnson90}.

This result was criticized by Magri {\it et al.} \cite{magri90} who a decade 
ago deduced from calculations for ordered compounds that the net charge of 
the atomic sphere of an alloy component is proportional to the number of 
nearest neighbors of the opposite type, and on this basis developed the 
charge correlated model in which the Madelung energy of a binary random 
alloy is given by

\begin{equation} \label{eq:E_CC}
E_{Mad}^{rand-ss} = -\frac{e^2}{2} \frac{\alpha_{cc}}{S} 
c(1-c)(q_A - q_B)^2 ,
\end{equation}
where $\alpha_{cc}$ is a constant equal to 0.5947049 and 0.60817846 for the 
fcc and bcc structures, respectively. Exactly the same result has been 
obtained later by Korzhavyi {\it et al.} \cite{korzhavyi92,abrikosov92} and 
by Johnson and Pinski \cite{johnson93} in their "screened" models for the
single-site CPA-DFT. The difference between these models lies only in the 
way the parameter $\alpha_{cc}$ is determined (from 0.4397212 for bcc in 
Ref.\ \cite{johnson93} to 0.54282038 for fcc in Ref.\ \cite{abrikosov98}). 
A discussion of the issues involved may be found in Refs.\ 
\cite{johnson93,korzhavyi95,ruban95,wolverton96,abrikosov98}.

It is certainly surprising, that (\ref{eq:E_CC}) is exactly the first term 
in (\ref{eq:E_Harr}) which was shown to vanish in the case of two charges,
here $q_A$ and $q_B$, randomly distributed on an underlying lattice in a 
compensating homogeneous effective medium. This means that either the models 
are inconsistent with general theory or that there should be some reason, not 
accounted for in the derivation by Krasko, for the presence of the first term 
in (\ref{eq:E_Harr}). 

Assuming the existence of only on-site and pair-wise interactions the 
Hamiltonian of a binary A$_c$B$_{1-c}$ alloy may be written

\begin{eqnarray} \label{eq:H_1}
H &=& \sum_{R} [ \epsilon^A_0 c_R + \epsilon^B_0 (1-c_R)] +
\frac{1}{2}\sum_{R \neq R'} [v^{AA}_{RR'}c_Rc_{R'} +
v^{AB}_{RR'}(1-c_R)c_{R'} \\ \nonumber
&+&  v^{AB}_{RR'}c_R(1-c_{R'}) +  v^{BB}_{RR'}(1-c_R)(1-c_{R'})] ,
\end{eqnarray}
where $\epsilon^{X}_{0}$ are on-site or intrasite interactions, which we will
assume depend only on the type of atom on site $R$, $v^{XY}_{RR'}$ are pair 
potentials acting between X and Y atoms at site $R$ and $R'$, respectively, 
and $c_R$ is the site-occupation operator taking on the value 1 if there is 
an A atom on site $R$ and 0 otherwise. Using $\delta c_R$, defined by 
$c_R = c + \delta c_R$, we may rewrite the Hamiltonian in the equivalent form

\begin{equation} \label{eq:H_2}
H = \frac{1}{2}\sum_{R,R'}  V_{RR'} \delta c_R \delta c_{R'} +
\frac{1}{2}\sum_{R \neq R'}[ c^2 v^{AB}_{RR'} + 2c(1-c) v^{AB}_{RR'} +
(1-c)^2 v^{BB}_{RR'}] ,
\end{equation}
where the first term includes the intrasite interaction ($R=R'$)

\begin{equation} \label{eq:VR0}
V_{R=0} = 2[\frac{1}{(1-c)}\epsilon^A_0 + \frac{1}{c} \epsilon^B_0]
\end{equation}
as well as the intersite interactions ($R \neq R'$)

\begin{equation} \label{V}
V_{RR'}= v^{AA}_{RR'} + v^{BB}_{RR'} - 2v^{AB}_{RR'} .
\end{equation}
Upon Fourier transformation of the first term we find

\begin{eqnarray} \label{eq:H_3}
H &=& \frac{N}{2\Omega_{BZ}}\int_{BZ} dq \; V(q) c_q c^*_q
+ \frac{1}{2}\sum_{R \neq R'} [
c^2 v^{AB}_{RR'} + 2c(1-c) v^{AB}_{RR'} +  (1-c)^2 v^{BB}_{RR'}],
\end{eqnarray}
where the second term is the average contribution to the energy due to pair 
interactions which in the case of direct Coulomb ion-ion interactions is 
$q_xq_y/|{\bf R} - {\bf R}'|$ combined with the corresponding contribution
from the interaction between the ions and the homogeneous compensating 
charge. This is exactly the second term in (\ref{eq:E_Harr}). 

The first term in (\ref{eq:H_3}) is usually associated with the 
configurational contribution to the energy of the system, but this is correct 
only if the contribution from intrasite interactions is zero. It is easily 
evaluated in a completely random alloy, where all the occupation numbers are
uncorrelated and therefore $c_q c^*_q = c(1-c)/N$ (which provides
the normalization of the ordering energy per atom). One finds

\begin{equation} \label{eq:1stterm}
\int_{BZ} dq \; V(q) c_q c^*_q = c(1-c) \int_{BZ} dq \;
V(q) = \Omega_{BZ} c(1-c) V_{R=0} ,
\end{equation}
which according to (\ref{eq:H_1}) is equal to 
$c \epsilon^A_0 + (1-c)\epsilon^B_0$.

It now remains to define the on-site interaction term $\epsilon^i_0$, which 
results from the interaction of the net charge $q_i$ in the alloy with the 
corresponding screening charge, in such a way that $\epsilon^i_0$ and the 
corresponding on-site Coulomb potential $V_i$ given by (\ref{eq:V}) are 
consistent within DFT, i.e., $V_i = \delta \epsilon^i_0 /\delta q_i$: 

\begin{equation} \label{eq:E_scr}
\epsilon^i_0 = -\frac{e^2}{2} \frac{\alpha_{scr}}{S} q_i^2 .
\end{equation}
Using this definition the first term in the Hamiltonian (\ref{eq:H_3})
may be written 

\begin{eqnarray} \label{eq:E_scr_rand}
E^{rand-ss}_{Mad} = E_{Mad}^{scr-ss} &=&
-\frac{e^2}{2} \frac{\alpha_{scr}}{S}[ c q_A^2 + (1-c) q_B^2] =
-\frac{e^2}{2} c(1-c) \frac{\alpha_{scr}}{S} (q_A - q_B)^2  \\ \nonumber
&\equiv&  c(1-c)V_{scr}(R=0),
\end{eqnarray}
which is exactly the first term in (\ref{eq:E_Harr}) and which, according to
the above derivation, arises from on-site or intrasite interactions such
as the screening interactions in metallic alloys.

It is now clear why the first term in (\ref{eq:E_Harr}) is absent in the
work by Krasko: His derivation is based on inter-site Coulomb interactions 
only. Thus, Krasko's result is valid in the absence of the screening 
intrasite interactions, which is the case, for instance, in the 
pseudopotential formalism within second order perturbation theory 
\cite{harrison66}. However, at the same time Harrison's result is in fact
correct under the more general assumptions needed in the SS-DFT-CPA
calculations.

\subsection{The configurational part of the Madelung energy and potential}

The previous section may seem trivial: First we define our inter- and 
intrasite interactions in real space, then we Fourier transform, and from 
the Fourier transform we return to the initially defined intrasite term.
However, in some formalisms, such as pseudopotential theory or the $S^{(2)}$ 
formalism \cite{gyorffy83,staunton94,pinski98x}, $V(q)$ is already defined 
and this may lead to problems with the correct definition of the 
configurational part of the total energy. The point is that the intrasite 
interactions {\it do not}\/ contribute to the {\it configurational}\/ part 
of the total energy which in real space may be written as (here, we do not 
consider the contribution from multi-site interactions) \cite{ruban97}

\begin{equation} \label{eq:H_conf_R}
H_{conf} = \frac{1}{2} \sum_{R \neq R'} V_{RR'} \delta c_R \delta c_{R'} .
\end{equation}
Therefore, if the configurational Hamiltonian is written in terms of $V(q)$,
e.g., in the concentration wave formalism, it must be corrected by the
subtraction of the corresponding intrasite interaction, i.e.,  

\begin{eqnarray} \label{eq:H_conf_q}
H_{conf} &=& \frac{1}{2\Omega_{BZ}}\int_{BZ} dq \; V(q) c_q c^*_q -
\frac{1}{2} c(1-c)V_{R=0} \\ \nonumber
&=& \frac{1}{2\Omega_{BZ}}\int_{BZ} dq \; [V(q) -  V_{R=0}] c_q c^*_q ,
\end{eqnarray}
where we have used the sum rule for the concentration wave density 
$c_q c^*_q$: $\int_{BZ} dq \; V(q) c_q c^*_q = \Omega_{BZ} c(1-c)$.

The subtraction of the intrasite term in (\ref{eq:H_conf_q}) is crucial for 
obtaining the correct ordering energy in pseudopotential theory and the 
$S^{(2)}$ formalism \cite{gyorffy83,staunton94,pinski98x} as well as for 
making the whole theory consistent. Let us, for instance, consider the 
Madelung energy of a binary completely ordered alloy with two non-equivalent 
sublattices. It is easy to show that its Madelung energy can be presented 
exactly in the form (\ref{eq:E_Harr}). For instance, the Madelung energy of 
the L1$_0$ ordered phase is

\begin{equation} \label{eq:E_Mad_L10}
E_{Mad}^{L1_0} = -\frac{e^2}{2} \frac{\alpha_{L1_0}}{S} c(1-c)(q_A - q_B)^2 +
\frac{e^2}{2} \frac{\alpha_{fcc}}{S} \tilde{q}^2 ,
\end{equation}
where the last term is zero in the ASA since $\tilde{q}$=0 as in the random 
alloy case, but now $\alpha_{L1_0}$ is a constant which appears due to the 
intersite Coulomb interactions. In the Appendix of Ref.\ \cite{landa91} it
is shown that, in fact, $\alpha_{L1_0}(q_A - q_B)^2/S$ is the Fourier 
transform of the effective direct electrostatic interaction at the 
corresponding superstructure vector $k_{L1_0} = 2\pi/a(100)$, i.e.,
$V_{es}(k_{L1_0}) = \alpha_{L1_0}(q_A - q_B)^2/S$.

On the other hand, $E_{Mad}^{L1_0}$ may also be found as the sum of the 
electrostatic energy of the completely random alloy, $E^{rand-ss}_{Mad}$,
given by (\ref{eq:E_scr_rand}) and the ordering energy, $\Delta U$:
$E_{Mad}^{L1_0} = E^{rand-ss}_{Mad} + \Delta U$. Since the Madelung energy 
of the ordered L1$_0$ alloy, $E_{Mad}^{L1_0}$ is uniquely defined in terms 
of the corresponding Madelung constant, which has nothing to do with the 
screening in the alloy, it is obvious that such a screening term must be 
present in the ordering energy, $\Delta U$, to compensate the screening 
contribution. 

Indeed, as shown in the Appendix of Ref.\ \cite{ruban97} the ordering energy
in the L1$_0$ structure can be written in the form as

\begin{equation} \label{U_ord}
\Delta U_{Mad} = \frac{1}{8} \eta^2 [V_{es}(k_{L1_0}) - V_{scr}(R=0)] ,
\end{equation}
from which it is easy to see that in the in the completely ordered state,
where the long range order parameter $\eta$=1, the last term in (\ref{U_ord})
is exactly the Madelung energy of random alloy at the stoichiometric 
composition ($c(1-c)$ = 1/4), and thus 
$E^{rand-ss}_{Mad} + \Delta U = 1/8 \alpha_{L1_0}(q_A - q_B)^2/S$.

This illustrates an important point: The ordering energy represented in  
reciprocal space in the concentration wave formalism must be corrected by the 
exclusion of the intrasite term, otherwise the theory will not be consistent.
Equation (\ref{eq:H_conf_q}) gives the correct definition of the ordering
energy considered more thoroughly in the Appendix of Ref.\ \cite{ruban97}. 
The intrasite interaction must also be subtracted when one considers the 
energy of SRO effects, and thus the correct Krivoglaz-Clapp-Moss expression 
must have $V(q) -  V_{R=0}$ instead of $V(q)$, which is exactly the case in 
Krivoglaz's derivation \cite{krivoglaz64}. Note, however, that this problem 
does not exist if the Krivoglaz-Clapp-Moss expression is used together with 
the so-called Onsager correction \cite{staunton94} under the condition that 
it is properly defined.

The reason, why it was possible to calculate $\alpha_{scr}$ on the basis of 
the {\it ordered}\/ structures, is the fact that, in an ordered binary alloy 
with only {\it two}\/ non-equivalent sublattices, one has an exact 
cancellation of the screening contribution to the Madelung energy and 
potential. This does not happen, however, in the general case of a supercell 
with $n>2$ non-equivalent sublattices. Here, the Madelung energy may written 
as sum of the contribution from the intrasite screening interactions 

\begin{eqnarray} \label{eq:E_intra_N}
E^{scr-sc}_{Mad} &=& \frac{e^2}{2N} \frac{\alpha_{scr}}{S}\sum_i q_i^2 =
\frac{e^2}{2} \frac{\alpha_{scr}}{S} [ c \frac{1}{N_A}\sum_{i=A} q_{iA}^2 +
(1-c) \frac{1}{N_B}\sum_{i=B} q_{iB}^2 ] \\ \nonumber
[&\neq& \frac{e^2}{2} \frac{\alpha_{scr}}{S} (c<q_A>^2 + (1-c) <q_B>^2)]  ,
\end{eqnarray}
where $N_A$ and $N_B$ are the number of A and B atoms, respectively, and 
the ordering energy due to the intersite interactions

\begin{equation} \label{eq:U_Mad}
\Delta U_{Mad} = \frac{e^2}{2 S} \sum_{i} \gamma_i(\alpha_{k_i} - \alpha_{scr})
\Delta q^2_{k_i} .
\end{equation}
Here, $\gamma_i$ is a normalizing coefficient, $\alpha_{k_i}$ a constant due
to the {\it bare}\/ electrostatic interactions between the net charges for 
the superstructure vector $k_i$ which may be calculated from the Madelung
constants $\alpha_M^{ij}$ of the corresponding supercell similar to the
$\alpha_{L1_0}$ considered above, and $\Delta q_{k_i}$ the difference
between the charges in the crests and in the troughs of the concentration
wave in the supercell.  In the case of a binary alloy with two 
non-equivalent sublattices, there is only one $k_i$  and
$\Delta q_{k_i}= (q_A - q_B)$.

If $\alpha_{k_i}$ depends only on the structure and describe the {\it bare}\/
electrostatic interaction between net charges, then $\Delta q_{k_i}$ 
"dresses" these interactions according to the real charge distribution in the
alloy (an equivalent description in real space in the charge correlated model
is given by Wolverton and Zunger \cite{wolverton95}, who also show that the 
Madelung energy of the random alloy has intrasite character). If the net 
charges in the supercell are screened (or uncorrelated) at distances less
than half the period of the concentration wave with wave-vector $k_i$ then 
$\Delta q_{k_i}$=0 and the corresponding contribution to the ordering energy 
vanishes. If the supercell includes only long-range concentration waves, the 
corresponding ordering contribution to the Madelung potential and energy 
becomes zero.

Let us finally mention the fact that the Madelung energy of a random alloy 
obtained in supercell calculations (\ref{eq:E_intra_N}) is not equal to
the Madelung energy in the single-site calculations, and thus it cannot 
be used to obtain $\alpha_{scr}$. The reason is simply that the Madelung 
energy is not a self-averaging quantity. However, the Madelung potential is, 
and it is clear that

\begin{equation}
<V_X>^{sc} =  \frac{e^2}{N_X} \frac{\alpha_{scr}}{S} \sum_{i=x} q_{iX} =
e^2 \frac{\alpha_{scr}}{S} <q_X> = V_X^{ss} ,
\end{equation}
which allows one to use (\ref{eq:alpha}) to obtained $\alpha_{scr}$ in
the supercell calculations and shows why $\alpha_{rand}$ is exactly
equal to $\alpha_{scr}$.

\subsection{Intersite screened Coulomb interactions}

Although the screened Coulomb interactions have an intrasite character, they
may contribute to the effective {\it pair intersite}\/ interactions of the 
kind obtained in the generalized perturbation method (GPM) 
\cite{ducastelle91,ducastelle76} because the screening charge is located 
on several of the coordination shells around each atom. This was, in fact,
already recognized by Ducastelle \cite{ducastelle91} who derived the 
contribution to the GPM potentials from the screened Coulomb interactions in 
the framework of the Hartree-Fock tight-binding CPA theory. 

The existence of an additional electrostatic term due to the screening is 
also consistent with Andersen's the force theorem \cite{mackintosh80}, which 
states that the change in the total energy of a system due to some 
perturbation to first order is given by the change in the sum of the 
one-electron energies obtained from frozen one-electron potentials plus the 
change of the electrostatic energy due to the perturbation. In fact, this 
latter contribution from the screened Coulomb interaction has been completely 
neglected in a number of first-principles calculations of GPM interactions 
\cite{gonis87,turchi88,drchal92,singh93x}. Here, we will therefore show how 
the screening contribution to the GPM potentials may be defined and obtained 
on the basis of the calculated spatial distribution of the screening charge.

\begin{figure}
\centerline{\psfig{figure=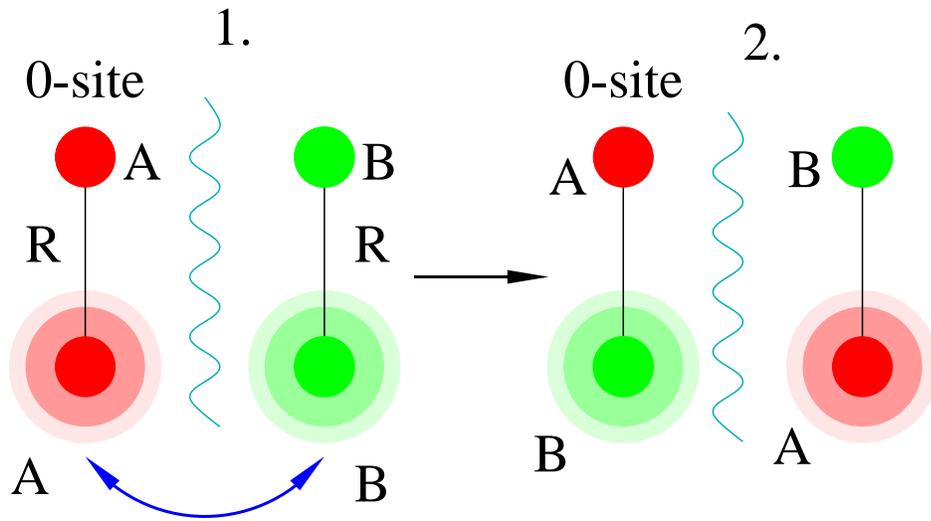,height=8.0cm}}
\caption[6]{Two systems, whose 0-site projected Coulomb energy is
to be used in the calculation of the screened effective interactions
at distance R.}
\label{Fig:Exch}
\end{figure}

GPM-like pair interactions, usually defined by (\ref{V}) for a specific
lattice vector ${\bf R}$, may be determined as the site-projected part of 
the change in the total energy when two atoms of different types in a 
completely random alloy are exchanged between sites infinitely far apart
in such a way that their neighbors at the relative position ${\bf R}$ are of 
the opposite type after the exchange. This is schematically illustrated in 
Fig.\ \ref{Fig:Exch}. That part of the total energy which should be accounted 
for is half the site-decomposed total energy written in terms of the 
intersite interactions or interatomic potentials, i.e.,

\begin{equation} \label{V2}
V^{(2)}(R) = \frac{1}{2} [E^{(2)}_1(R) - E^{(2)}_2(R) ] .
\end{equation}
Here, $E^{(1)}_1$ is the total energy due to pairwise interactions of the 
unperturbed system projected onto site $0$ and $ E^{(2)}_2$ is the same 
quantity after the exchange. A similar expression is also valid in the case 
of multisite interactions, but this will involve a more complex exchange of 
atoms and will not be considered here because the screened Coulomb 
interactions do not contribute to the effective multisite interactions in 
the ASA.  

Within multiple scattering theory as well as in the tight-binding 
approximation a Green's function formulation allows both site- and 
"path"-decomposition of the electron density and thereby makes it possible
to write down an analytical expression for the one-electron contribution
to the $n$-site interactions, $V^{(n)}(R)$, in the CPA \cite{ducastelle91,%
ducastelle76}. Concerning the screened Coulomb interactions one must, 
however, proceed differently. There are several ways to do so, but here we 
will present a straightforward approach.

In the sense of the CPA and single-site mean-field theory we will use an 
effective medium approach, assuming that at all sites, i.e., within the
atomic spheres assigned to each site, there is an electroneutral effective 
medium except at the two sites $\bf 0$ and $\bf R$ under consideration.  
In those two sites we must use the actual values of the net charges of the 
alloy components, which in the effective medium approach, are the average 
net charges, $q_A$ and $q_B$, of the alloy components. 

\begin{figure}
\centerline{\psfig{figure=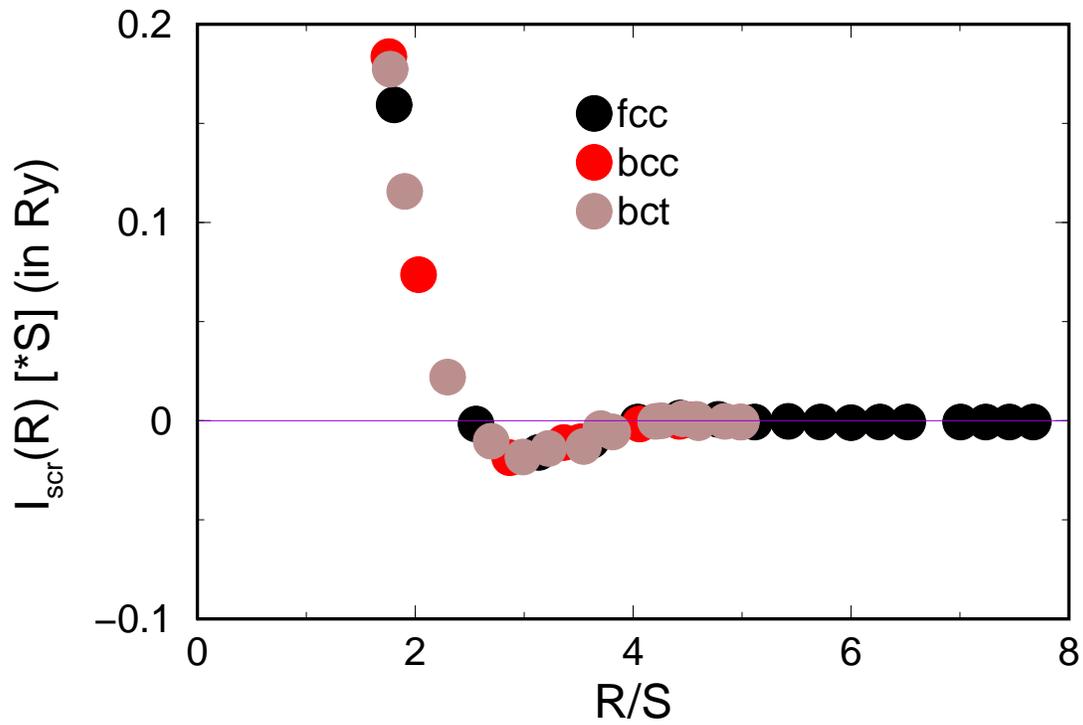,height=10.0cm}}
\caption[6]{The intersite screened Coulomb effective interactions
obtained from the normalized screening charge presented in Fig. 6.}
\label{Fig:VSCR}
\end{figure}

In the first-principles methods, however, these net charges depend on the
specific choice of the size of the atomic spheres and thus they must, 
in principle, go together with the corresponding screening cloud. Since we 
calculate the change in the electrostatic energy of the two systems shown in 
Fig.\ \ref{Fig:Exch} projected onto site $\bf 0$ due to the exchange of A 
and B atoms in positions ${\bf R}$, we must include only the interaction of 
the net charge at site $\bf 0$ with the net charge at site ${\bf R}$ and 
{\it its screening} charge. That is, the interaction of the net charge with 
its own screening charge must be excluded as it is included in the definition
of the screened on-site interactions, see (\ref{eq:alpha_scr}). Thus, the 
first term in (\ref{V2}) for the system before the exchange of atoms has 
been made, $E^{scr}_1(R)$, is

\begin{equation} \label{E_1}
E_1^{scr}(R) = e^2 q_A \sum_{{\bf R}' \neq 0} \frac{q_{A{\bf R}'}}{R'} + 
e^2 q_B \sum_{{\bf R}' \neq 0} \frac{q_{B{\bf R}'}}{R'} .
\end{equation}
Here, $q_{i{\bf R}'}$ is either the net charge of the $i$th component, 
$q_i$, if ${\bf R}'= {\bf R}$ or the corresponding screening charge if 
${\bf R}' \neq {\bf R}$. A similar expression may be written for 
$E^{scr}_2(R)$, after the exchange of the A and B atoms in the 
${\bf R}$-sites, i.e.,

\begin{equation} \label{E_2}
E_2^{scr}(R) = e^2 q_A \sum_{{\bf R}' \neq 0} \frac{q_{B{\bf R}'}}{R'} + 
e^2 q_B \sum_{{\bf R}' \neq 0} \frac{q_{A{\bf R}'}}{R'} .
\end{equation}
The resulting expression for the screened Coulomb interactions which should 
be added to the usual one-electron term is therefore 

\begin{eqnarray} \label{V_scr_GPM_1}
V_{scr}(R) &=& \frac{e^2}{2} [q_A \sum_{{\bf R}' \neq 0} 
\frac{q_{A{\bf R}'} - q_{B{\bf R}'}}{R'} - 
q_B \sum_{{\bf R}' \neq 0} \frac{q_{B{\bf R}'} - 
q_{A{\bf R}'}}{R'} ] \\ \nonumber
 &=& \frac{e^2}{2}(q_A-q_B)^2 \sum_{{\bf R}' \neq 0} 
\frac{Q(|{\bf R}'-{\bf R}|)}{R'} , 
\\ \nonumber 
\end{eqnarray}
where $Q(R)$ is the normalized screening charge defined in (\ref{Q_i}), and 
where we have used the condition that the screening does not depend on the
type of the atom. Finally, performing the summation in (\ref{V_scr_GPM_1}) 
one may define the screened Coulomb interactions as

\begin{equation} \label{V_scr(R)}
V_{scr}(R) = \frac{e^2}{2} (q_A-q_B)^2 \frac{\alpha_{scr}(R)}{S}.
\end{equation}
It is easy to see from (\ref{V_scr_GPM_1}) and (\ref{eq:alpha_scr}) that
${\alpha_{scr}(R=0)} = \alpha_{scr} =  \alpha_{rand}$ and therefore,
$V_{scr}(R=0)$ is exactly the on-site screened interactions that defines
the Madelung energy of the binary alloy which has exactly the same form
(\ref{eq:E_scr_rand}). This on-site interaction must be included in the 
definition of the $S^{(2)}$ interactions \cite{gyorffy83}, as has been 
demonstrated in the previous section (see also Ref.\ \cite{pinski98x}). 
When $R \neq 0$, $V_{scr}(R)$ defines the intersite screened Coulomb 
interaction contribution to the GPM-like effective interactions. Since 
the screening in the ASA, is practically universal these interactions 
have the universal form presented in Fig.\ \ref{Fig:VSCR}.

\section{The total energy in the single-site CPA and the supercell LSGF 
methods}

The fact that the Madelung energy of a random alloy described either by the 
effective medium model defined by the SS-DFT-CPA method or by the supercell 
model in conjunction with the SS-LSGF method differ from each other, has 
neither consequences for the final result for the total energy of the random 
alloy nor even for the partial and local contributions to the total energy. 
This follows simply from the fact that the density of states and its average 
local contributions are the same in the two methods, as shown above.

In Table \ref{E_iso_pol} we compare the total energy and its components in 
a Cu$_{50}$Pt$_{50}$ random alloy calculated by the SS-DFT-CPA method 
with $\alpha_{scr}=$0.60572 and by the SS-LSGF method on the basis 
of a 512-atom supercell, in which the atomic positions of Cu and Pt have
been chosen such that the SRO parameters are equal zero at the first 7
coordinations shells (LSGF-1). The agreement between the two calculations is
seen to be excellent if one combines the electron-nucleus, the 
electron-electron, and the Madelung contributions to form a total Coulomb 
energy, $E_{coul} = E_{el-nuc} + E_{el-el} + E_{Mad}$. 

The accuracy of the SS-DFT-CPA method with the appropriate screening
contribution to the Madelung potential and energy may be appreciated if one 
compares the results of a 512-atom supercell calculation performed by the 
SS-LSGF method (LSGF-2) where the distribution of the Cu and Pt atoms 
have not been optimized after the application of the random number generator 
leading to quite small, but not zero, SRO parameters. The values of the SRO 
parameters for the first 7 coordination shells are $-$0.005208 (1), 0.026041 
(2), 0.007161(3), -0.014323(4), -0.021484(5), 0.0390625(6), $-$0.0136718 (7),
respectively, which are approximately the same, as in the LSMS calculations
in Ref.\ \cite{faulkner97}. The agreement between SS-DFT-CPA results and
SS-LSGF calculations with a properly chosen supercell (LSGF-1) is obviously
better than between two SS-LSGF calculations.

\section{Conclusion}

The screened Coulomb interactions which are due to the interaction between 
the net charge of an alloy component and its screening charge must be 
included in a consistent single-site mean-field theory of the electrostatics 
in random alloys. In this paper we have shown how this may be done and we
have calculated the spatial distribution of the screening charge which in 
the ASA is found to be practically universal for homogeneous systems. A
formalism that describes the contribution from for the screened Coulomb  
interaction to Madelung potential and energy as well as to the effective 
interactions of the GPM-type is presented.

\section{Acknowledgements}

Valuable discussions with Dr.\ P.~A.\ Korzhavyi, I.~A. Abrikosov,
A.~Yu.\ Lozovoi, Prof.\ A.\ Gonis and Prof.\ S.\ Faulkner are greatly
acknowledged.  Center for Atomic-scale Materials Physics is sponsored
by the Danish National Research Foundation.

\noindent\begin{table}
\begin{tabular}{rrrrr}
Site &  Energy         &  ss-CPA-DFT   &     LSGF-1     & LSGF-2       \\
\hline
\hline
Cu &                  &                &                &                \\
   & Kinetic          & 3360.076110    &  3360.076294   & 3360.077674    \\
   & $<E_{el-nuc}>$   & $-$7974.160832 & $-$7974.157257 & $-$7974.178557 \\
   & $<E_{el-el}>$    & 1439.080777    &  1439.078359   & 1439.099272    \\
   & $<E_{Mad}>$      & $-$0.004193    &  $-$0.005646   &  $-$0.005994   \\
   & $<E_{Coul}>$     & $-$6535.084248 & $-$6535.084544 & $-$6535.085279 \\
   & $<E_{xc}>$       & $-$130.026085  & $-$130.026000  & $-$130.026621  \\
   & $<E_{Cu}>$       & $-$3305.034222 & $-$3305.034250 & $-$3305.034226 \\
\hline
Pt &                  &                 &                 &                 \\
   & Kinetic          & 42188.794140    & 42188.794273    & 42188.791806    \\
   & $<E_{el-nuc}>$   & $-$92747.101049 & $-$92747.093691 & $-$92747.030284 \\
   & $<E_{el-el}>$    & 14378.917863    &  14378.911671   & 14378.849707    \\
   & $<E_{Mad}>$      & $-$0.004193     & $-$0.005533     & $-$0.005472     \\
   & $<E_{Coul}>$     & $-$78368.187379 & $-$78368.187553 & $-$78368.186049 \\
   & $<E_{xc}>$       & $-$693.866856   & $-$693.866786   & $-$693.865919   \\
   & $<E_{Pt}>$       & $-$36873.260095 & $-$36873.260068 & $-$36873.260162 \\
\hline
\hline
Alloy & $E_{tot}$     & $-$20089.147159 & $-$20089.147159 & $-$20089.147194 \\
\hline
\end{tabular}
\caption[9]{
The total energy (in Ry) of Cu$_{50}$Pt$_{50}$ random alloy and corresponding 
contributions obtained in three different calculations: by the single-site 
CPA-DFT method (ss-CPA-DFT), in the 512-atom supercell LSGF calculations with 
optimized atomic distribution, providing zero SRO parameters up to the 7th 
coordination shell (LSGF-1), and with atomic configuration immediately after 
random number generator (LSGF-2). ($E_{coul} = E_{el-nuc}+ E_{el-el} + E_{Mad}$)
}
\label{E_iso_pol}
\end{table}

\end{document}